\documentclass[usenatbib,referee]{mn2e}
\usepackage{graphicx}
\usepackage{epsfig}
\usepackage{color}

\title[Evolving ONe WD+He star systems to IMBPs]
{Evolving ONe WD+He star systems to intermediate-mass binary pulsars}
\author[D. Liu et al.]
{D. Liu$^{\rm 1,2,3}$\thanks{E-mail:liudongdong@ynao.ac.cn}, B. Wang$^{\rm 1,2,3}$\thanks{E-mail:wangbo@ynao.ac.cn}, W. Chen$^{\rm 4}$, Z. Zuo$^{\rm 5}$ and Z. Han$^{\rm 1,2,3}$\\
$^1$Key Laboratory for the Structure and Evolution of Celestial Objects, Yunnan Observatories, Chinese Academy of Sciences,\\
 Kunming 650216, China\\
$^2$University of Chinese Academy of Sciences, Beijing 100049, China\\
$^3$Center for Astronomical Mega-Science, Chinese Academy of Sciences, Beijing 100012, China\\
$^4$School of Physics and Electrical Information, Shangqiu Normal University, Shangqiu 476000, China\\
$^5$School of Science, Xi'an Jiaotong University, Xi'an 710049, China}

\begin{document}
\date{}
\pagerange{\pageref{firstpage}--\pageref{lastpage}} \pubyear{2018}
\maketitle

\label{firstpage}

\begin{abstract}\label{0. abstract}
  It has been suggested that accretion-induced collapse (AIC) is a non-negligible path for the formation of the observed neutron stars (NSs). An ONe white dwarf (WD) that accretes material from a He star may experience AIC process and eventually produce intermediate-mass binary pulsars (IMBPs), named as the ONe WD$+$He star scenario. Note that previous studies can only account for part of the observed IMBPs with short orbital periods. In this work, we investigate the evolution of about 900 ONe WD$+$He star binaries to explore the distribution of IMBPs. We found that the ONe WD$+$He star scenario could form IMBPs including pulsars with $5-340\,\rm ms$ spin periods and $0.75-1.38\,\rm M_{\odot}$ WD companions, in which the orbital periods range from $0.04$ to $900\,\rm d$. Compared with the 20 observed IMBPs, this scenario can cover the parameters of 13 sources in the final orbital period$-$WD mass plane and the Corbet diagram, most of which has short orbital periods. We found that the ONe WD$+$He star scenario can explain almost all the observed IMBPs with short orbital periods. This work can well match the observed parameters of PSR J1802$-$2124 (one of the two precisely observed IMBPs), providing a possible evolutional path for its formation. We also speculate that the compact companion of HD 49798 (a hydrogen depleted sdO6 star) may be not a NS based on the present work.
\end{abstract}

\begin{keywords}
binaries: close -- stars: evolution -- supernovae: general -- white dwarfs -- stars: neutron 
\end{keywords}

\section{Introduction} \label{1. Introduction}
Neutron stars (NSs) are generally thought to be compact objects supported by nucleon-nucleon interaction via the strong force (e.g. Weber, Negreiros \& Rosenfield 2007).
There are three paths for the formation of NSs, i.e. the core-collapse supernovae of massive stars, the electron capture supernovae of intermediate-mass stars and the accretion-induced collapse (AIC) of massive white dwarfs (WDs; e.g. van den Heuvel 2009).
Although AIC events have not been directly observed, some recent studies indicate that at least some of the observed NSs may originate from the AIC channel, as follows:
(1) Hurley et al. (2010) found that the AIC channel cannot be ignored for the formation of millisecond pulsar (MSP) binaries, especially for the pulsar binaries with He WD companions and long orbital periods (see also Ivanova et al. 2008).
(2) Boyles et al. (2011) suggested that the observed pulsars in globular clusters, which have ages significantly less than the cluster ages, can be explained by the AIC channel that can produce newly born NSs with small kicks.
(3) In the strong magnetic fields, there exist some pulsars with He WD companions (e.g. Taam \& van den Heuvel 1986) and some accreting NSs in the low-mass X-ray binaries (LMXBs; e.g. van Paradijs et al. 1997; Xu \& Li 2009), which seem to have experienced extensive mass-transfer process but have not accumulate too much material onto the NS, and could be explained by recycling scenario of the AIC channel.
(4) The AIC channel can account for the large fraction of NSs remaining in the globular clusters and the formation of recycled pulsars with the low space velocities in the observations (e.g. Bailyn \& Grindlay 1990) owing to the small kick velocity and the small amount mass loss during the AIC process (e.g. Kitaura et al. 2006; Dessart et al. 2006).

There exist two progenitor models for AIC events, i.e. the single-degenerate (SD) model and the double-degenerate (DD) model. In the SD model, an ONe WD accretes H-rich matter from a main-sequence (MS) star or red-giant (RG) star, or alternatively accretes He-rich matter from a He star. The accreted matter will burn into O and Ne and accumulate onto the surface of the WD. When the ONe WD grows in mass close to the Chandrasekhar mass limit ($M_{\rm Ch}$), it collapses to a NS (e.g. Canal et al. 1990; Yungelson \& Livio 1998; Tauris et al. 2013; Ablimit et al. 2015; Brooks et al. 2017a). In the DD model, the merger of ONe WD$+$CO/ONe WD systems and double CO WDs may produce AIC events (e.g. Nomoto \& Iben 1985; Saio \& Nomoto 2004; Schwab, Quataert \& Kasen 2016).\footnote{The merging of double CO WDs may produce type Ia supernovae (SNe Ia) if the merging process is violent (e.g. Pakmor et al. 2010; Ruiter et al. 2013; Liu et al. 2016, 2017).} It is notable that the DD model seems quite difficult to produce MSPs or mildly recycled pulsars as the formed NSs will not experience recycled processes though some works have made attempts (e.g. Ransom et al. 2005).

Tauris et al. (2013) investigated the binary evolution of ONe WD$+$MS/RG/He star systems that may experience AIC and then be recycled to form binary pulsars. By considering the influence of irradiation-excited wind from the mass donors, Ablimit \& Li (2015) explored the evolution of ONe WD+MS systems, and found that their models can reproduce LMXBs with strong-field NSs and MSP$+$He WD systems with orbital periods less than a few days. The ONe WD$+$MS systems may evolve to form fully recycled MSPs with He WD companions, while the ONe WD$+$RG star systems are more likely to form intermediate-mass binary pulsars (IMBPs) including mildly recycled pulsars and CO/ONe WDs with long orbital periods. The present work focus on the ONe WD$+$He star systems, which may experience AIC and eventually form IMBPs with short orbital periods, named as the ONe WD$+$He star scenario.

It is generally thought that most of IMBPs evolve from intermediate-mass X-ray binaries (IMXBs) including an NS and a $2.0$$-$$10.0\,\rm M_{\odot}$ donor star (e.g. van den Heuvel 1975). However, this evolutionary channel cannot produce IMBPs with short orbital periods ($<$$3\,\rm d$; see Tauris, van den Heuvel \& Savonije 2000). Chen \& Liu (2013) suggested that the NS$+$He star evolutionary channel can explain the formation of 4 IMBPs with short orbital periods. Employing the anomalous magnetic braking of Ap/Bp stars, Liu \& Chen (2014) argued that IMXBs with Ap/Bp stars can form some IMBPs with short orbital periods (see also Justham, Rappaport \& Podsiadlowski 2006; Shao \& Li 2012).

However, these previous studies can only account for part of IMBPs with short orbital periods (e.g. Chen \& Liu 2013).
Although the ONe WD$+$He star scenario has been investigated by Tauris et al. (2013), they only considered the case with the initial mass of the primary ONe WDs $M^{\rm i}_{\rm WD}=1.2\,\rm M_{\odot}$.
In order to explore the possibility of the ONe WD$+$He star scenario in forming IMBPs detailedly, we evolve a large number of ONe WD$+$He star systems for the formation of recycled pulsars with CO/ONe WD companions with $M^{\rm i}_{\rm WD}=1.0, 1.1, 1.2$ and $1.3\,\rm M_{\odot}$. Compared with Tauris et al. (2013), we found that the initial regions of ONe WD$+$He star systems for producing AIC in the present work are significantly enlarged, mainly owing to different assumptions about the behavior at high mass-transfer rates. We then present the parameter space of the NS$+$He star systems just after AIC, and provide the parameter space of the finally formed IMBPs. We found that almost all the observed IMBPs with short orbital periods can be covered by the ONe WD$+$He star scenario. We also reproduce the current stage of a precisely observed IMBP PSR J0621+1002 in our calculations.

This paper is organized as follows.
In Sect.\,2, we present the methods for the evolution of ONe WD$+$He star systems that can experience AIC and eventually form IMBPs. The corresponding results are provided in Sect.\,3. We gave a discussion in Sect.\,4 and finally a summary in Sect.\,5.

\section{Numerical code and methods} \label{2. Methods}
We employ the Eggleton stellar evolution code that has been updated for the last decades (e.g. Eggleton 1973; Han, Podsiadlowski \& Eggleton 1994; Pols et al. 1995, 1998; Eggleton \& Kiseleva-Eggleton 2002) to simulate the evolution of ONe WD$+$He star systems. We adopt the He mass fraction $Y$$=$$0.98$ and the metallicity $Z$$=$$0.02$ for the initial He models. We also consider the orbital angular momentum loss derived from the gravitational wave radiation (see Landau \& Lifshitz 1971). The mass loss during the mass-growth process are assumed to take away specific orbital angular momentum of the compact accretors.

In order to calculate the mass-transfer process, we use the prescription described in Han, Tout \& Eggleton (2000). According to this prescription, the donor will overflow its Roche-lobe stably and as necessary, but never too much. During the mass-transfer process, the prescriptions for the mass-growth rate of the WD are similar to that described in Wang et al. (2009). If the mass-transfer rate ($\dot{M}_{\rm 2}$) is higher than a critical mass-transfer rate $\dot{M}_{\rm cr}$ provided in Nomoto (1982), the binary is assumed to enter an optically thick wind process (Hachisu, Kato \& Nomoto 1996). In this stage, the accreted He-rich matter is supposed to burn stably, and accumulate onto the surface of the WD at the rate of $\dot{M}_{\rm cr}$; the rest of He-rich matter is assumed to be blown away from the binary system in the form of optically thick wind. 
If the mass-transfer rate is lower than the critical rate and higher than the minimum accretion rate for stable He burning $\dot{M}_{\rm st}$, we assume that the He burns stably and no mass is lost from the system (see Wang et al. 2009). If the mass-transfer rate is lower than $\dot{M}_{\rm st}$, the accreted He would experience He-flash process and accumulate onto the surface of the WD. In this case, the mass-growth rate of the WDs is defined as
\begin{equation}
\dot{M}_{\rm WD}=\eta_{\rm He}\dot{M}_{\rm 2},
\end{equation}
where $\eta_{\rm He}$ is the mass-accumulation efficiency for He-shell flash process (see Kato \& Hachisu 2004).

When the WD grows in mass to $M_{\rm Ch}$ (set to be $1.38\,\rm M_{\odot}$), we assume that the WD would experience AIC process and become a NS. During this process, a mass equivalent of $0.13\,\rm M_{\odot}$ of the WD is supposed to convert into the released gravitational binding energy, forming a NS with the gravitational mass of $1.25\,\rm M_{\odot}$ after the AIC process (e.g. Ablimit \& Li 2015). The orbital separation turns to be wider due to the sudde mass loss. We adopt the prescription of angular momentum conservation to calculate the relationship between the orbital separations just before and after AIC ($a_{\rm 0}$ and $a$; see Verbunt, Wijers \& Burn 1990), written as
\begin{equation}
\frac{a}{a_{\rm 0}}=\frac{M_{\rm WD}+M_{\rm 2}}{M_{\rm NS}+M_{\rm 2}},
\end{equation}
where the $M_{\rm WD}$, $M_{\rm 2}$ and $M_{\rm NS}$ are the mass of the WD just before AIC, the He star and the formed NS just after AIC, respectively. Here, the binary orbit is assumed to be re-circularized after AIC. In our calculations, we do not include the influence of kick velocity, since the influence of a kick velocity with a dispersion of $50\,\rm km/s$ may be not serious based on the calculations of Hurley et al. (2010) and Tauris et al. (2013). We also discuss the influence of the kick velocity on our final results in Sect.\,4.4.

After the AIC process, the He star may fill its Roche-lobe again, and transfer He-rich matter and angular momentum onto the formed NS, during which the binary can be detected as a LMXB or an IMXB (e.g. Podsiadlowski, Rappaport \& Pfahl 2002). We employed the prescription described in Tauris et al. (2013) to calculate the mass-growth rate of the NS ($\dot{M}_{\rm NS}$), written as
\begin{equation}
\dot{M}_{\rm NS}=(|\dot{M}_{\rm 2}|-\max[|\dot{M}_{\rm 2}|-\dot{M}_{\rm Edd},0])\cdot e_{\rm acc}\cdot k_{\rm def},
\end{equation}
in which $\dot{M}_{\rm Edd}$ is the Eddington mass-accretion rate, $e_{\rm acc}$ is the fraction of the transferred material from the He star that is actually accreted and remains on the NS, and $k_{\rm def}$ is the ratio of gravitational mass to the rest mass of the accreted material. We combined $e_{\rm acc}$ and $k_{\rm def}$ as a single free parameter (i.e. $e_{\rm acc}k_{\rm def}$) to describe the mass-growth process of NSs, and set $e_{\rm acc}\cdot k_{\rm def}=0.35$ (see also Ablimit \& Li 2015). However, the value of $e_{\rm acc}\cdot k_{\rm def}$ is still uncertain, which is set by the evidence of the inefficient accretion for LMXBs (e.g. Jacoby et al. 2005; Antoniadis et al. 2012). According to Eq.\,(3), when $\dot{M}_{\rm 2}$ is larger than $\dot{M}_{\rm Edd}$, $\dot{M}_{\rm NS}=0.35\dot{M}_{\rm Edd}$; when $\dot{M}_{\rm 2}$ is lower than $\dot{M}_{\rm Edd}$, $\dot{M}_{\rm NS}=0.35\dot{M}_{\rm 2}$. Here, $\dot{M}_{\rm Edd}$ is written as
\begin{equation}
\dot{M}_{\rm Edd}=2.3\times 10^{\rm -8}\,M_{\odot}\,\rm yr^{\rm -1}\cdot M^{\rm -1/3}_{\rm NS}\cdot \frac{2}{1+X},
\end{equation}
in which $X$ is the fraction of H in the mass donor that is set to be zero for He stars.
In addition, the accreted material of the NS ($\Delta M_{\rm NS}$) may recycle the formed NS. We assume that the initial spin angular momentum of the formed NS from AIC is negligible before the accretion process, and adopt the relationship between $\Delta M_{\rm NS}$ and the spin period of the recycled NS ($P_{\rm spin}$) presented in Tauris, Langer \& Kramer (2012), written as
\begin{equation}
P_{\rm spin}\approx 0.34\times(\Delta M_{\rm NS}/M_{\odot})^{\rm -3/4},
\end{equation}
where $P_{\rm spin}$ is in units of ms.
This prescription is obtained by the assumption that the release of gravitational binding energy of the accreted matter is negligible.

\section{Results} \label{3. Results}
\subsection{A typical example for binary evolution}
\begin{figure*}
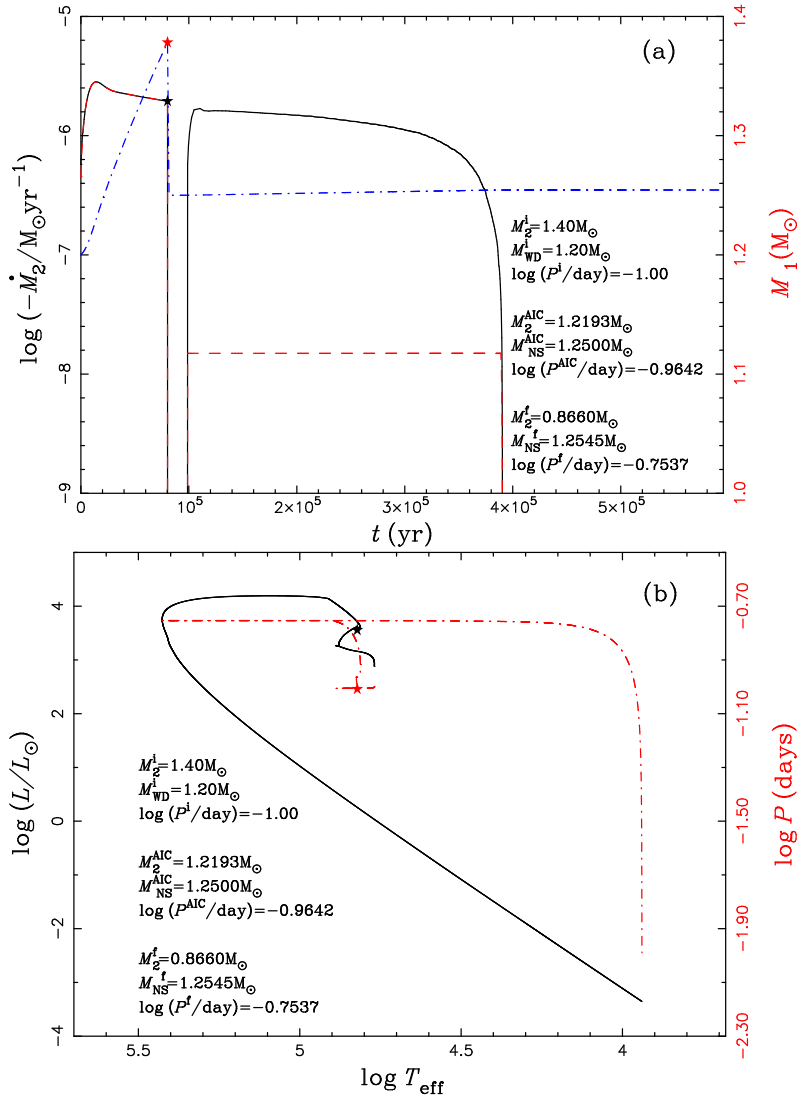

\epsfig{file=f1a.ps,angle=270,width=10.5cm}\ \
\epsfig{file=f1b.ps,angle=270,width=10.5cm}
  \caption{A typical example for the evolution of an ONe WD$+$He star system until the formation of an IMBP.  Panel (a) presents the evolution of the $\dot{M}_{\rm 2}$ (the black solid curve), $\dot{M}_{\rm WD}$ (or $\dot{M}_{\rm NS}$; the red dashed curve) and $M_{\rm WD}$ (or $M_{\rm NS}$; the blue dash-dotted curve) as a function of time. Note that the light curve for $\dot{M}_{\rm 2}$ and $\dot{M}_{\rm WD}$ in this panel are overlapped for the pre-AIC evolution as $\dot{M}_{\rm 2}$ is in the stable He burning range.
  Panel (b) shows the luminosity of the mass donor (the black solid curve) and the binary orbital
  period (the red dash-dotted curve) as a function of effective temperature.
  The filled stars in the panels (a) and (b) represent the position where AIC occurs.
  The initial binary parameters of WD$+$He star system, the binary parameters of NS$+$He star systems just after AIC, and the eventually formed NS$+$WD systems at its formation moment are also given in these two panels.}
\end{figure*}

Fig.\,1 shows the evolution of an ONe WD$+$He star system that experiences AIC process and eventually forms an IMBP. The initial binary has a $1.2\,\rm M_{\odot}$ WD and a $1.4\,\rm M_{\odot}$ He star with an orbital period of $0.1\,\rm d$. The He star expands and fills its Roche-lobe when it evolves to the He subgiant stage. After that, the binary quickly enters the stable He-shell burning stage, during which the He-rich matter burns stably and no mass is lost from the system. After about $8.3\times10^{\rm 4}\,\rm yr$, an AIC event occurs when the WD grows in mass to $M_{\rm Ch}$ and the orbital period is $0.098\,\rm d$. After the AIC process, the binary turns to have a $1.25\,\rm M_{\odot}$ NS and a $1.2193\,\rm M_{\odot}$ He star with an orbital period of $0.1086\,\rm d$. At this time, the He star becomes below its Roche-lobe as the orbital separation turns to be longer after the sudden mass loss, which has been used to explain the formation of redbacks (e.g. Smedley et al. 2015). After about $1.7\times10^{\rm 4}\,\rm yr$, the He star refills its Roche-lobe and starts to transfer He-rich matter onto the NS, leading to a recycled process for the NS. After that, the mass-transfer rate quickly increase to be significantly higher than the Eddington accretion rate. In this case, part of the accreted mass is blown away from the binary by the radiation pressure of the NS, and the NS grows in mass slowly at the rate of $0.35\dot{M}_{\rm Edd}$. About $2.93\times10^{\rm 5}\,\rm yr$ later, the He star exhausts its He-shell and evolves to a CO WD. The binary eventually evolves to an IMBP consisting a $1.2545\,\rm M_{\odot}$ Pulsar with a spin period of $19.57\,\rm ms$ and a $0.8660\,\rm M_{\odot}$ CO WD. At this moment, the orbital period of the IMBP is about $0.1763\,\rm d$, and this binary will merge in about $1.1\times10^{\rm 9}\,\rm yr$.

\subsection{Parameter spaces}
\begin{figure}
\begin{center}
\epsfig{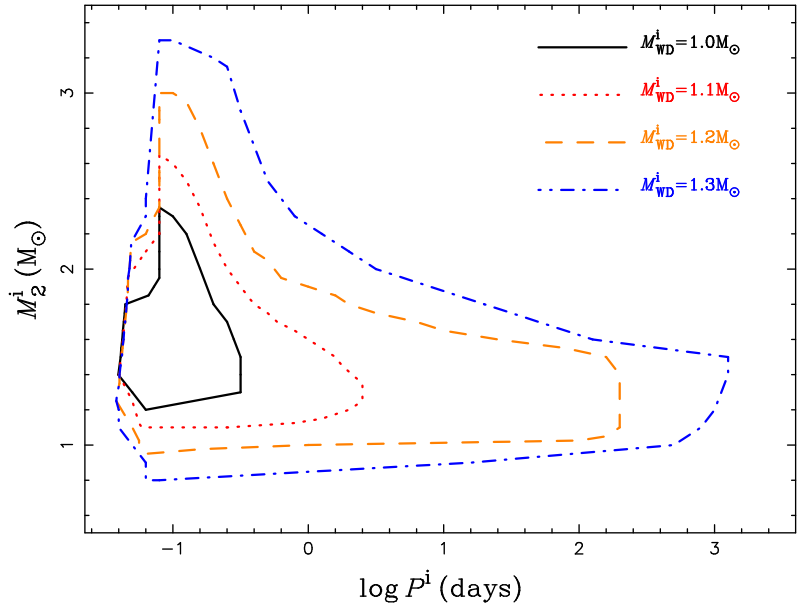}
 \caption{Initial parameter space of ONe WD$+$He star systems that can experience AIC in the initial orbital period$-$initial secondary mass ($\log P^{\rm i}$$-$$M_{\rm 2}^{\rm i}$) plane. The solid, dotted, dashed and dash-dotted curves represent the cases with initial WD mass $M_{\rm WD}=1.0, 1.1, 1.2$ and 1.3, respectively.}
  \end{center}
\end{figure}

We have evolved more than 900 WD$+$He star binaries and thus obtained the parameter space for these binaries in different evolutionary phases.
Fig.\,2 presents the initial parameter space of WD$+$He star systems that can produce AIC events in the initial orbital period$-$secondary mass ($\log P^{\rm i}$$-$$M_{\rm 2}^{\rm i}$) plane. The initial masses of the WDs ($M^{\rm i}_{\rm WD}$) are in the range of $1.0-1.3\,\rm M_{\odot}$ for different contours. In order to form AIC events, the binaries should have He stars with initial masses of $0.8-3.3\,\rm M_{\odot}$ and orbital periods of $\sim$$0.04$$-$$1300\,\rm d$.
The binary systems outside of the contours in Fig.\,2 will not experience AIC events. The systems beyond the upper boundaries will experience mass-transfer process at a relatively high rate owning to the large mass-ratio, which may result in CE processes. The ONe WDs in the systems beyond the lower boundaries cannot grow in mass to the $M_{\rm Ch}$ due to the low $\dot{M}_{\rm 2}$ and the low $M_{\rm 2}^{\rm i}$.
The left boundaries are determined by the conditions that the He stars fill their Roche-lobe when they are on the He zero-age MS stage. The systems beyond the right boundaries will undergo rapid mass-transfer process owning to the rapid expansion of the He stars, leading to the mass-loss of too much material in the form of the optically thick wind. As a result, the binaries beyond the upper, right and lower boundaries will eventually form double WD systems.

\begin{figure}
\begin{center}
\epsfig{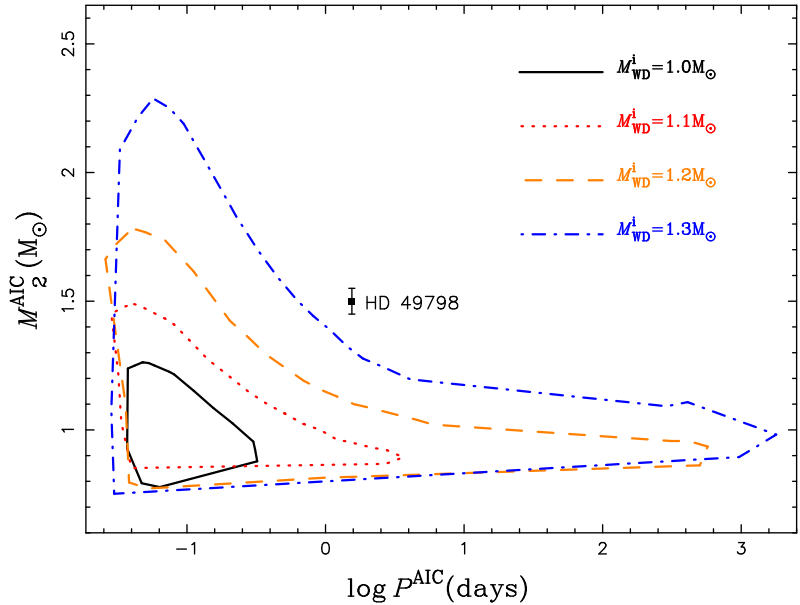}
 \caption{Similar to Fig.\,2, but for the parameter space of NS$+$He star systems just after the AIC processes. The filled squared with error bar represents a H depleted subdwarf O6 star HD 49798 with its compact companion RX J0648.0$-$4418 (see Thackeray 1970; Mereghetti et al. 2009).}
  \end{center}
\end{figure}

When the WD in the WD$+$He star systems grows in mass close to $M_{\rm Ch}$, it may experience AIC process and form a NS. Fig.\,3 shows the space parameters of the formed NS$+$He star systems just after the AIC processes in the orbital period$-$secondary mass ($\log P^{\rm AIC}$$-$$M_{\rm 2}^{\rm AIC}$) plane. We found that the He stars have masses in the range of $0.75-2.28\,\rm M_{\odot}$, and the orbital periods are from $\sim$$0.04$ to $1800\,\rm d$ at this moment.

\begin{table*}
\begin{center}
 \caption{The parameters of 20 observed IMBPs taken from the ATNF Pulsar Catalogue in 2017 October (Manchester et al. 2005; see http://www.atnf.csiro.au/research/pulsar/psrcat). The WD masses marked by $\ast$ are obtained from timing, and others are the median masses that are calculated by assuming the orbital inclination angle $i=60^{\circ}$ and the pulsar mass $M_{\rm NS}=1.35\,\rm M_{\odot}$ (filled stars in Fig.\,4). Note that the WD masses will be slightly lower the values given here when we assume $M_{\rm NS}=1.25\,\rm M_{\odot}$.}
   \begin{tabular}{ccccccccc}
\hline \hline
 $\rm No.$ & $\rm Pulsars$ & $\rm P_{\rm spin}/ms$ & $\rm P_{\rm orb}/d$ & $\rm M_{\rm WD}/M_{\odot}$\\
\hline
$1$ & $\rm J0621+1002$ & $28.9$ & $8.32$ & $0.67^{*}$\\
$2$ & $\rm B0655+64$ & $195.7$ & $1.03$ & $0.80$\\
$3$ & $\rm J1022+1001$ & $16.5$ & $7.81$ & $0.85$\\
$4$ & $\rm J1141-6545$ & $393.9$ & $0.20$ & $1.02^{*}$\\
$5$ & $\rm J1157-5112$ & $43.6$ & $3.51$ & $1.30^{*}$\\
$6$ & $\rm J1244-6359$ & $147.3$ & $17.17$ & $0.69$\\
$7$ & $\rm J1337-6423$ & $9.4$ & $4.79$ & $0.95$\\
$8$ & $\rm J1435-6100$ & $9.3$ & $1.35$ & $1.08$\\
$9$ & $\rm 1439-5501$ & $28.6$ & $2.12$ & $1.30^{*}$\\
$10$ & $\rm J1454-5846$ & $45.2$ & $12.4$ & $1.05$\\
$11$ & $\rm J1525-5545$ & $11.4$ & $0.99$ & $0.99$\\
$12$ & $\rm J1528-3146$ & $60.8$ & $3.18$ & $1.15$\\
$13$ & $\rm J1727-2946$ & $27.1$ & $40.3$ & $1.01$\\
$14$ & $\rm J1757-5322$ & $8.9$ & $0.45$ & $0.67$\\
$15$ & $\rm J1802-2124$ & $12.6$ & $0.70$ & $\textbf{0.78}^{*}$\\
$16$ & $\rm J1933+1726$ & $21.5$ & $5.15$ & $0.94$\\
$17$ & $\rm J1949+3106$ & $13.1$ & $1.95$ & $0.97$\\
$18$ & $\rm J1952+2630$ & $20.7$ & $0.39$ & $1.13$\\
$19$ & $\rm J2222-0137$ & $32.8$ & $2.5$ & $1.38$\\
$20$ & $\rm J2303+46$ & $1066$ & $12.3$ & $1.30^{*}$\\
\hline \label{1}
\end{tabular}
\end{center}
\end{table*}

\begin{figure}
\begin{center}
\epsfig{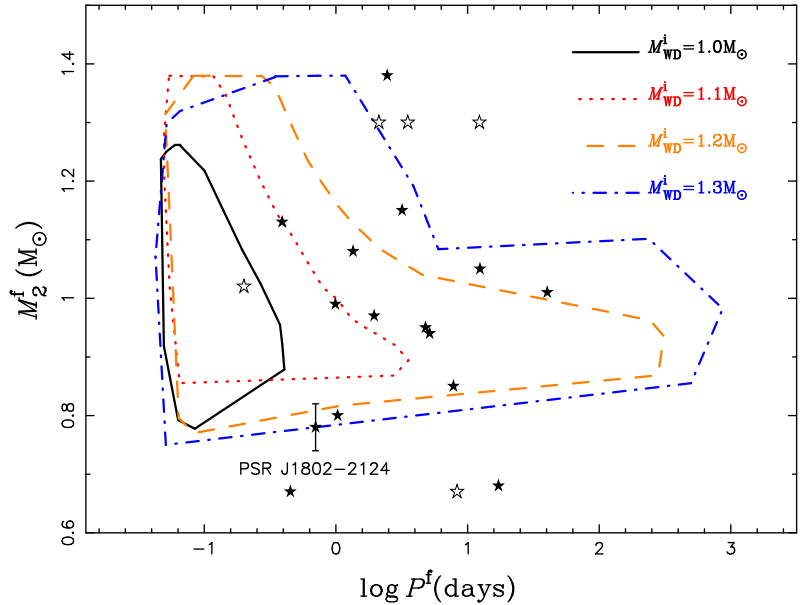}
 \caption{Similar to Fig.\,2, but for the parameter space of the formed IMBPs at their formation moment. The star symbols represent 20 observed IMBPs listed in Table\,1 (see Manchester et al. 2005). The open and filled stars represent the WD masses that are obtained from timing and median masses, respectively. The filled star with error bar represents a precisely measured IMBP, PSR J1802$-$2124 (Ferdman et al. 2010).}
  \end{center}
\end{figure}

In Fig.\,4, we present the distributions of the finally formed NS$+$hot WD systems in the final orbital period$-$ secondary mass ($\log P^{\rm f}$$-$$M_{\rm 2}^{\rm f}$) plane. The WDs in the IMBPs here are just formed hot CO/ONe WDs. For the predicted IMBPs, the orbital periods are distributed between $\sim$$0.04\,\rm d$ and $900\,\rm d$, and the mass of the WDs range from $0.75$ to $1.38\,\rm M_{\odot}$. The IMBPs with shorter orbital periods can also be produced through the ONe WD$+$He star scenario after the gravitational wave radiation. The 20 observed IMBPs with CO/ONe WD companions listed in Table 1 are plotted by stars. From this figure, we can see that 13 IMBPs are located in the predicted parameter space of NS$+$WD systems from our calculations, which indicates that the AIC channel of ONe WD+He star systems is an important path for the formation of IMBPs.
Note that some He stars may still have masses larger than $M_{\rm Ch}$ after the recycled process. These super-$M_{\rm ch}$ He stars may explode as type Ib supernovae and collapse to be NSs (e.g. Yoon et al. 2010, 2012; Zhu, L\"u \& Wang 2015). In this case, double NSs including mildly recycled pulsars are produced, which may be good candidates for gravitational wave sources. In the present work, we only considered the formation of NS$+$WD system (i.e. IMBPs). Further investigations of the formation of double NSs from this scenario are needed.

\begin{figure}
\begin{center}
\epsfig{file=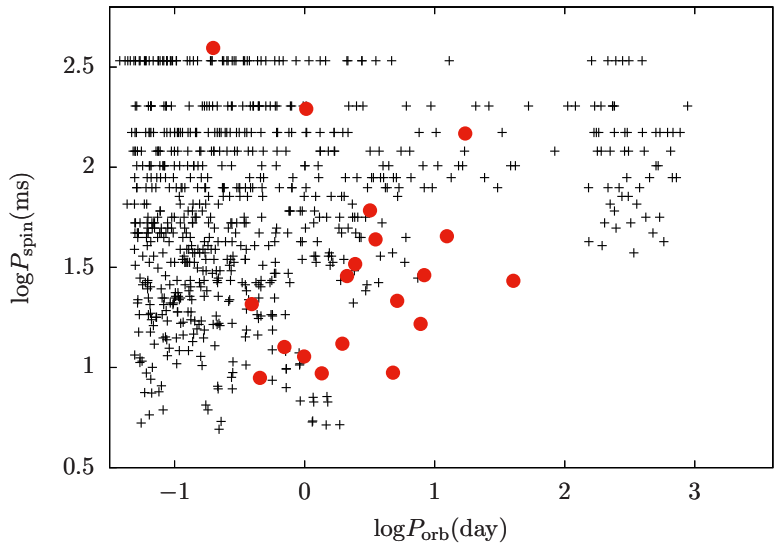,angle=0,width=11.5cm}
 \caption{Parameter space of the eventually formed IMBPs in the Corbet diagram based on the ONe WD$+$He star scenario. The orbital periods are measured at the formation moment of IMBPs. The filled circles represent the observed IMBPs listed in Table\,1 (see Manchester et al. 2005).}
  \end{center}
\end{figure}

The NSs formed through the AIC channel may be recycled after the mass donors refill their Roche-lobe. Fig.\,5 shows the parameter space of our predicted IMBPs in the orbital period$-$spin period ($\log P_{\rm orb}$$-$$P_{\rm spin}$) plane (i.e. the Corbet diagram; Corbet 1984). In this figure, the spin periods of NSs are in the range of about $5-340\,\rm ms$ based on the ONe WD$+$He star scenario. Note that some NSs are almost not spun-up (i.e. $\Delta M_{\rm NS}<10^{\rm -4}\,\rm M_{\odot}$) in our calculations, which are not plotted here. From this figure, we can see that the spin periods take discrete values. This is because the NS masses are reserved to only five valid digits and the cases with the same $\Delta M_{\rm NS}$ result in the same spin periods (see Eq.\,5). In this figure, we also plot the observed IMBPs listed in Table\,1. We found that the orbital and spin periods of 13 observed IMBPs can be covered by this scenario. Note that previous studies can only account for part of the observed IMBPs with short orbital periods. In the present work, we found that almost all the observed IMBPs with short orbital periods can be covered. Note that PSR J2303$+$46 is not plotted in this figure, which belongs to the non-recycled NSs (see Tauris \& Sennels 2000).

\subsection{PSR J1802$-$2124} \label{3. PSR J1802-2124}
PSR J1802$-$2124 is one of the two well observed IMBPs whose pulsar masses have been precisely measured. It was first observed by the Parkes Multibeam Pulsar Survey (Faulkner et al. 2004). PSR J1802$-$2124 has a $1.24\pm0.11\,\rm M_{\odot}$ NS and a $0.78\pm0.04\,\rm M_{\odot}$ CO WD companion with an orbital period of $0.7\,\rm d$ and a spin period of $12.6\,\rm ms$ (Ferdman et al. 2010). Fig.\,6 shows a possible path for the formation of PSR J1802$-$2124, in which the initial binary has a $1.3\,\rm M_{\odot}$ ONe WD and a $1.0\,\rm M_{\odot}$ He star with an orbital period of $0.47\,\rm d$. The subsequent mass-transfer process is similar to that presented in Fig.\,1, and the binary experiences an AIC process forming a NS$+$He star system. Afterwards, the He star refills its Roche-lobe and finally produce an IMBP system, which includes a $1.2583\,\rm M_{\odot}$ NS and a $0.8079\,\rm M_{\odot}$ CO WD with an orbital period of $0.7\,\rm d$ at its formation moment. In order to reproduce the spin period of PSR J1802$-$2124, we assume the coefficient $e_{\rm acc}\cdot k_{\rm def}=1.0$ during the mass-transfer process from the He star onto the NS (such a large value of $e_{\rm acc}\cdot k_{\rm def}=1.0$ is also widely used, e.g. Chen, Li \& Xu 2011), and obtain a spin period of $\sim$$12\,\rm ms$. At this moment, the observed parameters of PSR J1802$-$2124 are well reproduced.
Note that Chen, Li \& Xu (2011) have proposed that a binary including a $1.3\,\rm M_{\odot}$ NS and a $1.0\,\rm M_{\odot}$ He star with an orbital period of $0.5\,\rm d$ may also reproduce the current parameters of PSR J1802$-$2124. They found that the eventually formed IMBP for reproducing PSR J1802$-$2124 has a spin period of $<16\,\rm ms$ and an orbital period of $0.71\,\rm d$. However, the present work matches the observed spin period and orbital period of PSR J1802$-$2124 better.

\begin{figure*}
\epsfig{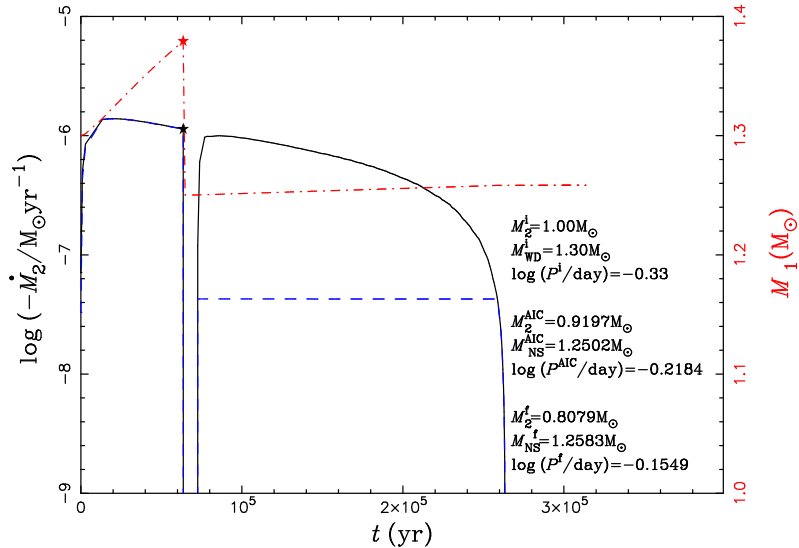}
\caption{A possible evolutionary path for the formation of PSR J1802$-$2124. Here, the coefficient $e_{\rm acc}\cdot k_{\rm def}$ is assumed to be 1.0. The filled stars represent the positions where AIC occurs.}
\end{figure*}

\section{Discussions} \label{4. Discussion}
\subsection{Compared with previous studies}
There are some other formation channels for IMBPs, e.g. the IMXB evolutionary channel and the NS$+$He star channel. (1) In the IMXB evolutionary channel, a $1.5-10\,\rm M_{\odot}$ H-rich donor star transfers material onto the surface of the NS, leading to the spun-up of the NS (van den Heuvel 1975). When the H-rich donor star evolves to a CO/ONe WD, an IMBP system is produced (e.g. Podsiadlowski, Rappaport \& pfahl 2002; Tauris, Langer \& Kramer 2012). However, this channel cannot produce IMBPs with orbital periods less than $3\,\rm d$ as the orbital separation would be widen due to the mass-transfer from the less massive donors to the more massive NSs once the IMXBs evolve to LMXBs (see Tauris et al. 2000). (2) In the NS$+$He star channel, the NSs are produced from the iron core-collapse supernovae and electron-capture supernovae. The NSs are recycled by the accretion process from the He star, and the binary evolves to an IMBP when the He star exhausts its He-rich shell (e.g. Chen \& Liu 2013). Chen \& Liu (2013) argued that this channel cannot reproduce all IMBPs with short orbital periods. In the present work, we found that the ONe WD$+$He star scenario can produce IMBPs with wide orbital periods ranging from 0.04 to $900\,\rm d$, and almost all the observed IMBPs with short orbital periods can be covered by this scenario (see Fig.\,5).

In addition, Tauris et al. (2013) provided a parameter space of ONe WD$+$He star systems for producing AIC events with an initial mass of WD $M^{\rm i}_{\rm WD}=1.2\,\rm M_{\odot}$. Their work shows that the initial masses of He stars are from 1.1 to $1.5\,\rm M_{\odot}$ and the initial orbital periods are in the range of $0.04$$-$$1.2\,\rm d$. Compared with the results of Tauris et al. (2013), the initial parameter space in the present work obviously has more massive He stars and longer orbital periods. This is because Tauris et al. (2013) assumed that the binaries will evolve to common envelope process when $\dot{M}_{\rm 2}$ is larger than the critical mass-transfer rate $3\dot{M}_{\rm Edd,WD}$. We note that $\dot{M}_{\rm Edd,WD}$ they adopted is for H-accreting WDs, but too small for He-accreting WDs as the opacity of He is lower than H. Tauris et al. (2013) also pointed out that the value of $\dot{M}_{\rm CE}$ may be the largest uncertainty in their calculations. In addition, Tauris et al. (2013) adopted $M_{\rm Ch}=1.48\,\rm M_{\odot}$ for rigidly rotating WDs, which could also shrink their initial parameter space for AIC events.

\subsection{HD 49798/RX J0648.0$-$4418}
HD 49798/RX J0648.0$-$4418 is a binary including a $1.5\,\rm M_{\odot}$ H depleted subdwarf O6 star and a $1.28\,\rm M_{\odot}$ X-ray pulsating companion with an orbital period of $1.548\,\rm d$ (e.g. Thackeray 1970; Stickland \& Lloyd 1994; Mereghetti et al. 2009). The compact companion RX J0648.0$-$4418 is generally thought to be a WD (e.g. Mereghetti et al. 2011). It has been suggested that HD 49798/RX J0648.0$-$4418 is a good candidate for the progenitor of SNe Ia (e.g. Mereghetti et al. 2009, 2013; Wang \& Han 2010b; Liu et al. 2015). Recently, Mereghetti et al. (2016) argued that the spinning up rate of the pulsating companion is best interpreted in terms of a NS accreting from the wind of the subdwarf HD 49798 (see also Brooks, Kupfer \& Bildsten 2017). However, the large emitting radius ($R\sim40~\rm km$) derived from the black body spectral remains puzzling.

In Fig.\,3, we compared HD 49798/RX J0648.0$-$4418 (the filled square with error bar) with the form NS$+$He star systems from the AIC channel, which indicates that a NS$+$He star system with parameters like HD 49798/RX J0648.0$-$4418 seems unlikely to be formed from the AIC process of an ONe WD$+$He star system. Chen, Li \& Xu (2011) presented the distribution of NS$+$He star systems originating from iron core-collapse supernovae or electron capture supernovae, which still cannot cover the parameters of HD 49798/RX J0648.0$-$4418. Thus, we speculate that the companion of HD 49798 may be a WD but not a NS. Recently, Popov et al. (2018) also suggested that the spin-up of the compact object in this source could originate from the contraction of a young WD.

\subsection{AIC or SNe Ia}
Recently, it has been suggested that a CO WD$+$He star system may also produce AIC events (e.g. Brooks et al. 2016; Wang, Podsiadlowski \& Han 2017). Wang, Podsiadlowski \& Han (2017) recently suggested that a CO WD that accretes matter from a He star may trigger off-centre carbon ignition if the accretion rate is larger than a critical value ($\sim$$2.05\times10^{\rm -6}M_{\odot}\,\rm yr^{\rm -1}$), leading to an AIC event rather than an SN Ia. These AIC events can also produce NS$+$He star systems, in which NSs may also be recycled when the He stars refill their Roche-lobe, leading to the formation of IMBPs finally. Thus, further investigations of the properties of IMBPs from the AIC processes of CO WD$+$He star systems are needed.

In addition, Marquardt et al. (2015) argued that ONe WDs may explode as an SN Ia though they do not demonstrate a mechanism that would cause the ONe WDs to explode (see also Miyaji et al. 1980). They performed hydrodynamic explosion simulations for the detonations in initially hydrostatic ONe WDs, and concluded that the properties of ONe WD thermonuclear explosions are quite similar to SNe Ia that originate from CO WDs with similar masses. They also argued that about 3$-$10\% of potential progenitor systems for SNe Ia should contain ONe WDs. Jones et al. (2016) simulated the O deflagration processes in the degenerate ONe cores, and argued that a thermonuclear explosion might be produced (see also Isern, Canal \& Labay 1991).
However, recent simulations indicate that the accreting ONe WDs actually result in the AIC events and finally produce NSs (e.g. Schwab, Quataert \& Bildsten 2015; Brooks et al. 2017a; Wu \& Wang 2018).

\subsection{Uncertainties}
Our results may be influenced by some factors, e.g. the metallicity, the kick velocity, the accretion disks, the effect of evaporation, the rotation, the behavior at high mass-transfer rates, etc.
(1) The initial parameter space of ONe WD$+$He star systems may be changed if a different metallicity is adopted; the initial parameter space would be enlarged to have larger He star masses and longer orbital periods for a higher metallicity (e.g. Wang \& Han 2010a), leading to a wider distributions of the final produced IMBPs.
(2) Tauris et al. (2013) have investigated the influence of kick velocity on the formed IMBPs from the ONe WD$+$He star scenario. According to the He5, He6 and He7 models in Tauris et al. (2013), a larger kick velocity would result in the formation of an IMBP with larger orbital periods (about 1.5 and 2 times for the cases with the kick velocity of $50\,\rm km/s$ and $450\,\rm km/s$). Thus, the observed IMBPs with orbital periods near $\sim$$10\,\rm d$ that are located outside of our parameter space in Figs\,4 and 5 may be covered by the ONe WD$+$He star scenario when the kick velocity is considered.
(3) In some IXBPs, the thermal-viscous instability in the accretion disks may reduce the active X-ray time and the amount of mass-accumulated onto the NSs, resulting in the production of more mildly recycled pulsars (Li 2002).
(4) When a pulsar is formed in a binary system, it may evaporate the envelope of its companion with its high-energy radiation/particles if no mass-transfer interaction occurs (e.g. van den Heuvel \& van Paradijs 1988; Ruderman et al. 1989). Liu \& Li (2017) investigated the effect of evaporation both before and during the Roche-lobe overflow process, and found that this effect may influence the initial spin period and surface magnetic field of the newborn NSs.
(5) The rotation of ONe WDs may increase their critical mass for producing AIC events, which will shrink the initial parameter space presented in Fig.\,2 (e.g. Ablimit \& Li 2015).
(6) As mentioned in Sect. 4.1, the behavior of WD$+$He star systems is still under debate at super-Eddington accretion rates, which may have significant influence on the initial parameter space for AIC events (see Tauris et al. 2013).

\section{Summary} \label{5. Summary}
In this work, we investigate the ONe WD$+$He star scenario for the production of IMBPs. We summarize our main results as follows:

(1) In order to produce AIC events, the ONe WD$+$He star system should have He stars with initial masses of $0.8$$-$$3.3\,\rm M_{\odot}$ and initial orbital periods of $0.04$$-$$1300\,\rm d$ (see Fig.\,2).

(2) Just after the AIC process, the masses of He stars are in the range of $0.75$$-$$2.28\,\rm M_{\odot}$, and the orbital periods range from 0.04 to $1800\,\rm d$ (see Fig.\,3).

(3) According to the ONe WD$+$He star scenario, the formed IMBPs contain $0.75$$-$$1.38\,\rm M_{\odot}$ CO/ONe WDs and pulsars with $5$$-$$340\,\rm ms$ spin periods, in which the orbital periods are in the range of $0.04$$-$$900\,\rm d$. For the 20 observed IMBPs, the parameters of 13 sources can be covered by this scenario, in which all the IMBPs with short orbital periods are covered. However, our evolutionary route is difficult to form IMBPs with long orbital period, which may arise from IMXB evolutionary channel (see Figs\,4 and 5).

(4) The current parameters of PSR J1802$-$2124 can be well matched by the ONe WD$+$He star scenario when we set the combined coefficient $e_{\rm acc}k_{\rm def}=1.0$ for the mass-transfer process in NS$+$He star systems, in which the initial binary has a $1.3\,\rm M_{\odot}$ ONe WD and a $1.0\,\rm M_{\odot}$ He star with an orbital period of $0.47\,\rm d$ (see Fig.\,6).

(5) We speculate that the X-ray pulsating companion of HD 49798 may be a WD but not a NS, since the calculations on the formation of NS$+$He stars do not produce binaries with the observed parameters of HD 49798/RX J0648.0$-$4418 (see Fig.\,4).

\section*{Acknowledgments}
We acknowledge useful comments and suggestions from the anonymous referee.
We also thank Philipp Podsiadlowski for his helpful discussions. 
This study is supported by the 973 Program of China (No. 2014CB845700), the Natural Science Foundation of China (Nos. 11673059, 11521303, 11573021, 11573016 and 11390374), Chinese Academy of Sciences (Nos. KJZD-EW-M06-01 and QYZDB-SSW-SYS001), the Fundamental Research Funds for the Central Universities, the Program for Innovative Research Team (in Science and Technology) at the University of Henan Province, the Natural Science Basic Research Program of Shaanxi Province$-$Youth Talent Project (No. 2016JQ1016) and Yunnan Province (Nos. 2013HA005, 2013HB097 and 2017HC018).

\label{lastpage}
\end{document}